\documentclass[aps,prl,reprint,superscriptaddress,twocolumn,showpacs,amsmath,amssymb]{revtex4-1}
\usepackage{ulem}

\usepackage{graphicx}
\usepackage{amsmath}
\usepackage{color}
\usepackage{graphicx}

\begin{document}
\newcommand{\ud}{{\mathrm d}}
\newcommand{\sech}{\mathrm{sech}}
\newcommand{\sinc}{\mathrm{sinc}}
\newcommand{\cor}[1]{\color{red}{#1}}

\title{Irrationality and quasiperiodicity in driven nonlinear systems}

\author{David Cubero}
\email[]{dcubero@us.es}
\affiliation{Departamento de F\'{\i}sica Aplicada I, EUP, Universidad de Sevilla, Calle Virgen de \'Africa 7, 41011 Sevilla, Spain}
\author{Jes\'us Casado-Pascual}
\email[]{jcasado@us.es}
\affiliation{F\'{\i}sica Te\'orica, Universidad de Sevilla, Apartado de Correos 1065, 41080 Sevilla, Spain}
\author{Ferruccio Renzoni}
\email[]{f.renzoni@ucl.ac.uk}
\affiliation{Department of Physics and Astronomy, University College London, Gower Street, London WC1E 6BT, United Kingdom}

\date{\today}

\begin{abstract}
We analyse the relationship between irrationality 
and quasiperiodicity in nonlinear driven systems. To that purpose we consider a nonlinear system whose steady-state response is very sensitive to the periodic or quasiperiodic character of the input signal. In the infinite time limit, an input signal consisting of two incommensurate frequencies will be recognised by the system as quasiperiodic. We show that this is in general not true in the case of finite  interaction times. An irrational ratio of the driving frequencies of the input signal is not sufficient for it to be recognised by the nonlinear system as quasiperiodic,  resulting in observations which may differ by several orders of magnitude from the expected quasiperiodic behavior. Thus, the system response depends on the nature of the irrational ratio, as well as the observation time. We derive a condition for the input signal to be identified by the system as quasiperiodic. Such a condition also takes into account the sub-Fourier response of the nonlinear system. 
 
\end{abstract}


\maketitle

\paragraph*{Introduction.--} Periodic structures, in space and in time, are ubiquitous in all branches of science, and the characteristics of many systems can be traced back to their periodicity. Periodic systems can be defined by rational numbers. In the case of spatially periodic systems, the potential is the sum of harmonics with lattice constants in rational ratio, while in 
the time-periodic case the driving is made of harmonics with frequencies in rational ratio (commensurate frequencies).  Quasiperiodic systems are obtained whenever the ratio between spatial or temporal harmonics is irrational (incommensurate frequencies). Quasiperiodic order has been observed in solid state structures \cite{levine} and cold atom systems \cite{guidoni}. Experiments and simulations on driven nonlinear systems revealed  a number of distinguishing features associated with the quasiperiodic nature of the driving  \cite{feudel06, neupik02, gomden06, cubren12}.

In a real experiment, quasiperiodic structures can be  represented by their best periodic approximations. This raises the issue of whether the system dynamics corresponds to the original quasiperiodic structure, or to the approximant periodic one. This is an essential issue whenever periodicity and quasiperiodicity lead to completely different dynamics.
A standard approach \cite{glalib88} 
relies on the ``most irrational numbers'', the {\it golden ratio} being the most popular choice,  as determined by using Farey tree type classification \cite{sos58,sla67}, or equivalently the continued fraction representation of the number.  
The use of these numbers guarantees that their best rational approximations are the best choice for 
the system response to reflect the original quasiperiodicity of the temporal or spatial structure. 
However this leaves open the important question of what level of irrationality is required for the system to react following a genuine quasiperiodic behavior given the finite dimensions of the experiment. This is precisely the question addressed here.

In this work we consider a nonlinear driven system whose response in the infinite time limit is very sensitive to the quasiperiodic nature of the driving, i.e. the system reacts in a completely different way depending on whether the driving is periodic or quasiperiodic. We study the response of the system  to a bi-harmonic drive in the finite interaction time limit. We first show analytically that the frequency resolution of the system is sub-Fourier and provide an expression for it. Then  we examine the response of the system to a drive consisting of incommensurate frequencies.  We show that irrationality alone is not sufficient to guarantee quasiperiodic behavior. Instead, the response of the system depends on the nature of the irrational frequency ratio and on the interaction time. 

\paragraph*{ Model system, periodic driving, and  sub-Fourier resolution.--}
As a case study of a system that is very sensitive to  whether the drive is periodic or quasiperiodic,  we consider a  driven classical particle. 
The system  dynamics is described, in the deterministic and overdamped regime, by 
\begin{equation}
\gamma\dot{x}(t)=-U'[x(t)]+F(t),
\label{eq:overdamp}
\end{equation}
where the dot and the prime denote time and spatial derivatives, respectively, $\gamma$ is the friction coefficient, $U(x)=U_0\cos(2kx)/2$ is a spatially periodic and symmetric potential, with period $\lambda=\pi/k$, 
and $F(t)$ is a driving force. 
Note however that the conclusions reported in this paper are based only on symmetry considerations, and, thus, do not depend on the specific details of the dynamics \footnote{See Supplemental Material at  for the explicit validation of our approach in the case of a spatially non-periodic system.}.

The system response to the drive can be characterised by the average velocity
$v=\lim_{T_s\rightarrow\infty}(x(T_s)-x(0))/T_s$.
In any spatially symmetric system, 
 $v$  is reversed when the driving is inverted, 
\begin{equation}
v[-F(t)]=-v[F(t)].
\label{eq:spatsymm}
\end{equation}
Further, it must not depend on the specific choice of the time origin, $v[F(t+t_0)]=v[F(t)]$. By combining both transformations, it follows that the current vanishes if the driving possesses the following symmetry $F_{sh}:~ F(t+t_0)=-F(t)$,
for all $t$ and a given value $t_0$. If the system is also overdamped, like the one described by (\ref{eq:overdamp}), another symmetry that must be broken for a finite current is given by \cite{rei01,reimann02}
$F_s:~ F(t_0-t)=-F(t_0+t)$.
A common choice used to break both symmetries is the bi-harmonic driving $F_1(t)=F_0[ \cos(\omega_1 t)+\cos(\omega_2 t) ]  $,
with two commensurate frequencies $\omega_1$ and $\omega_2$, i.e. $\omega_2/\omega_1=p/q$ with $p$ and $q$ being two co-prime positive integers. Another driving that will be used here to produce a current is $F_2(t)=F_0[\cos(\omega_1t)+\cos(2\omega_1t)]\cos(\omega_2t)$.
These driving forces are particular examples of the generic driving  $F(t)=\Phi(\omega_1 t+\varphi_1,\omega_2 t+\varphi_2),  $
where $\Phi$ is a function that is periodic in both its arguments, $\Phi(x_1+2\pi,x_2)=\Phi(x_1,x_2+2\pi)=\Phi(x_1,x_2)$, and $\varphi_1$ and $\varphi_2$ are constants. The driving period itself is given by 
 $T=2\pi q/\omega_1=2\pi p/\omega_2$.
The current invariance under the choice of the time origin implies invariance under the transformation $\varphi_1\rightarrow \varphi_1+\omega_1 t_0$, $\varphi_2\rightarrow \varphi_2+\omega_2 t_0$. By choosing $t_0=-\varphi_1/\omega_1$, it follows that the current $v=v_{pq}$ depends on the driving phases only through the combination $\theta=q\varphi_2-p\varphi_1$. On the other hand, using B\'ezout's  lemma it is easy to show that the periodicity of $\Phi(x_1,x_2)$ on both its arguments implies 
\begin{equation}
v_{pq}(\theta+2\pi)=v_{pq}(\theta).
\label{eq:vpq_periodic}
\end{equation}
This last symmetry property, together with (\ref{eq:spatsymm}), can be readily used to show that if $\Phi$ satisfies any of the following conditions: (i) $\Phi(x_1+\pi,x_2+\pi)=-\Phi(x_1,x_2)$, (ii)  $\Phi(x_1+\pi,x_2)=-\Phi(x_1,x_2)$, or (iii)  $\Phi(x_1,x_2+\pi)=-\Phi(x_1,x_2)$, then
\begin{equation}
v_{pq}(\theta+\pi)=-v_{pq}(\theta).
\label{eq:vpq_shiftsym}
\end{equation}
From (\ref{eq:vpq_shiftsym}) it readily follows that when $p$ and $q$ are both odd and (i) holds ---as is the case for the driving $F_1$--- the symmetry 
 $F_{sh}$ holds and thus $v_{pq}(\theta)=0$ for all $\theta$. Similarly, the current vanishes when $p$ is even in case (ii), and when $q$ is even in case (iii) ---the driving $F_2$ fulfils (iii).

When both symmetries  $F_{sh}$ and $F_{s}$ are broken in a deterministic overdamped system, a finite current of the form  $v_{pq}=(m/n) v_0$ is expected \cite{ajdmuk94}, where $m$ and $n$ are two integers, and $v_0=\lambda/T$. 

The response of the system to a quasiperiodic driving is completely different  \cite{neupik02,fladen04,gomden06,cubren12}. It can be shown, on the basis of general symmetry properties only, that there is no current when the two driving frequencies $\omega_1$ and $\omega_2$ are incommensurate \cite{neupik02,cubren12}.

\begin{figure}
\begin{center}
\includegraphics[width=0.35\textwidth]{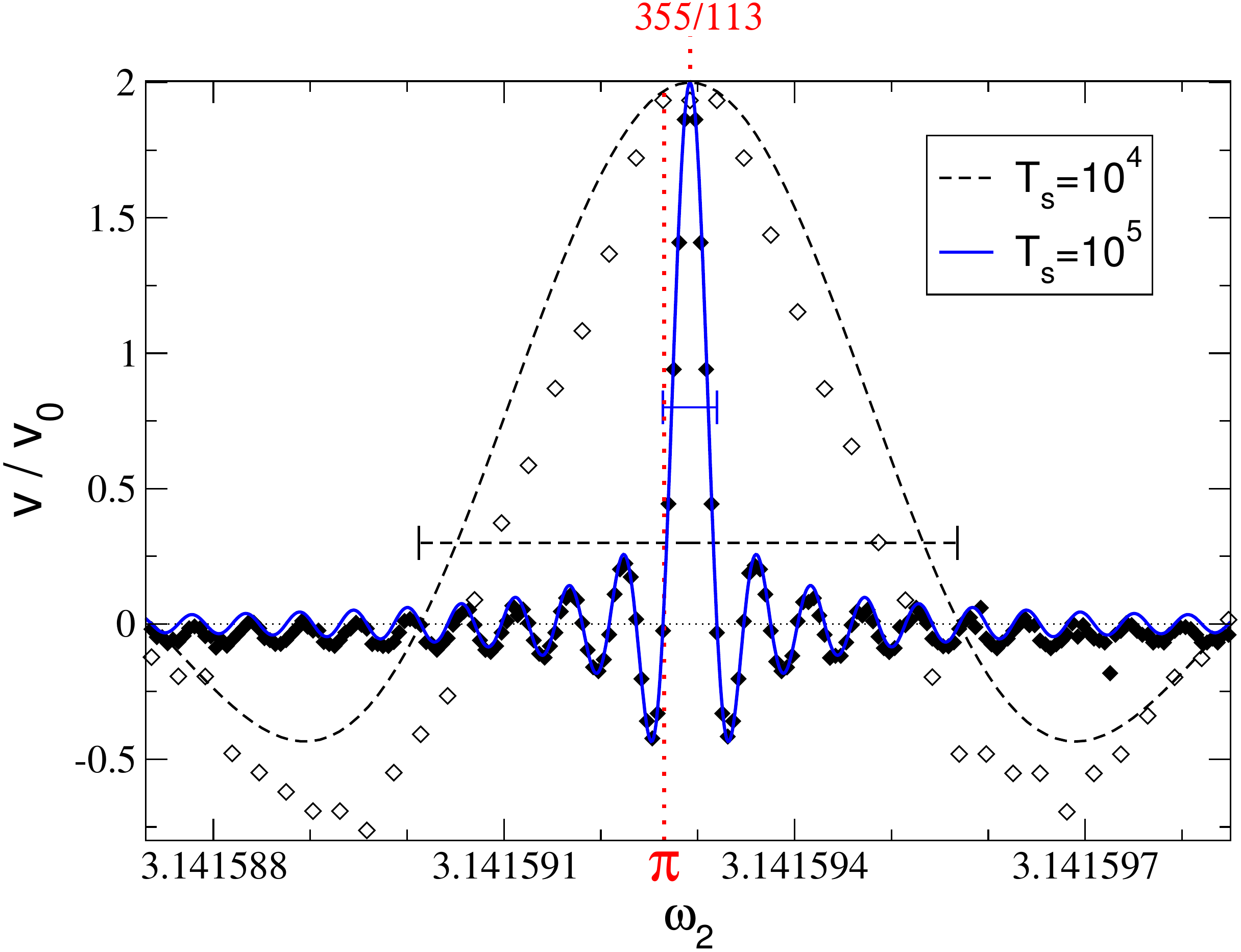}
\end{center}
\caption{Current  vs. driving frequency  $\omega_2$ for the overdamped system (\ref{eq:overdamp}) with the driving $F_2$ and $F_0=5.75$. Reduced units are defined in the simulations such that $U_0=k=\gamma=\omega_1=1$. Empty and filled diamonds correspond to $T_s=10^4$ and $10^5$, respectively.   The lines are the predictions given by (\ref{eq:v:Ts:overdamp}) with $q=113$,  $p=355$, and $v_0=1/(2q)$. The horizontal bars depict the frequency width (\ref{eq:freqres}), showing a resolution 113 times smaller than that expected from the Fourier width $2\pi/T_s$.}
\label{fig:freq_resolution} 
\end{figure}

The above results assume an infinite sampling time $T_s$. For large but finite times,  the following  expression for the finite-time current $v_{T_s}=(x(T_s)-x(0))/T_s$
\begin{equation}
v_{T_s}\sim  \frac{1}{(q\omega_2-p\omega_1)T_s}\int_\theta^{\theta+(q\omega_2-p\omega_1)T_s}\!\!\! d\theta^\prime \, v_{pq}(\theta^\prime)
\label{eq:v:Ts}
\end{equation}
was shown in \cite{cascub13} to describe the asymptotic evolution of the system, when the second driving frequency $\omega_2$ is in the neighborhood of $\omega_1 p/q$,  independent of the system details.
In overdamped systems, the leading expression obtained from a functional expansion on the driving  \cite{quicue10,cuequi13} is given by $v_{pq}=B\cos(\theta)$, which when inserted in (\ref{eq:v:Ts}) for $F_1$ and $F_2$ yields
\begin{equation}
 v_{T_s}\sim v_{pq} \frac{\sin[(q \omega_2-p\omega_1 )T_s]}{(q\omega_2 - p \omega_1)T_s}.
\label{eq:v:Ts:overdamp}
\end{equation}

Our numerical simulations presented in Fig. \ref{fig:freq_resolution}  are in good agreement  with (\ref{eq:v:Ts:overdamp}) for the overdamped system (\ref{eq:overdamp}), especially for larger $T_s$, when the current is more independent of the initial conditions. 

A general expression for the width of the resonance can be derived on the basis of symmetries only \cite{cascub13}, i.e. from Eqs. (\ref{eq:vpq_periodic}) and (\ref{eq:vpq_shiftsym}).
The integral in (\ref{eq:v:Ts}) cancels whenever $(q\omega_2-p\omega_1)T_s$ is a multiple of $2\pi$.  Furthermore,   the current, as a function of $\omega_2=\omega_1p/q+\delta\omega_2$,  decays as $(q\delta \omega_2 T_s)^{-1}$  as we move away from $\omega_2=\omega_1p/q$.  The distance to the first zero thus gives an estimation of the width of such a  resonance as
\begin{equation}
\Delta\omega_2=\frac{2\pi}{q\,T_s}.
\label{eq:freqres}
\end{equation}
The frequency window given by (\ref{eq:freqres}) is a factor $q$  smaller than the  Fourier width $\Delta \omega_F=2\pi/T_s$.
Numerical calculations for the specific model (\ref{eq:overdamp}) confirm the sub-Fourier width of the resonance, as shown  in Fig.~\ref{fig:freq_resolution}.

\paragraph*{Periodic approximations and  quasiperiodicity vs. irrationality.--}

The infinite-time current $v_{pq}$ associated with a periodic driving with $\omega_2=\omega_1 p/q$ can be accurately computed in a simulation or experiment provided the observation time $T_s$ is much larger than the driving period $T_q=2\pi q/\omega_1$. 

In turn, given an arbitrary frequency $\omega_2$, we may look for periodic drivings $\omega_1 p/q$ which provide a good approximation to the current. 
This will occur whenever both frequencies are close enough, that is, 
\begin{equation}
|\omega_2-\omega_1 p/q|<\varepsilon_0\Delta\omega_2,
\label{eq:crit}
\end{equation}
where $\varepsilon_0$ is a small dimensionless parameter. For example, by choosing $\varepsilon_0=0.1$, (\ref{eq:v:Ts:overdamp}) predicts current deviations from $v_{pq}$ smaller than 2\%.

Given the values $\omega_2$ and $T_s$, we can define the best periodic approximation from the lowest positive integers $p$ and $q$ such that (\ref{eq:crit}) is satisfied. The associated period $T_q$ can be used to predict the periodic or quasiperiodic behavior of the system under the given driving. If $T_q\ll T_s$, then periodic behavior is expected. Genuine quasiperiodicity would require $T_q\gtrsim T_s$ regardless of the specific value of $T_s$. As we discuss below, this does not occur for all irrational ratios $\omega_2/\omega_1$.

Equation (\ref{eq:crit}) can be rewritten as $\phi\le(T_q/T_s) \varepsilon_0 $,
where $\phi=|\omega_2/\omega_1-p/q|\,q^2$
is a function which measures how well the ratio $\omega_2/\omega_1$ is approximated by the rational approximation $p/q$. 
For a given approximation $p/q$,  this inequality 
provides a range of  times $T_s$ where the corresponding
periodic approximation is expected to hold, with the maximum value $T_s^\mathrm{max}$ given by 
\begin{equation} 
T_q = \frac{\phi}{\varepsilon_0} T_s^\mathrm{max}.
\label{eq:Tses}
\end{equation}
Therefore, even if $\omega_2/\omega_1$ is irrational, if there exists a rational approximation such that as $\phi/\varepsilon_0\ll1$, then there will be a certain range of observation times where $T_q\ll T_s$, and thus 
periodic, instead of quasiperiodic behavior, is expected. 

 Our analysis identifies the error function $\phi$ as the central quantity to determine the relationship between irrationality and quasiperiodicity in a physical system. Standard results of number theory,
stated in the following, will clarify how $\phi$ is related to different irrational numbers.

\paragraph*{Approximation of irrationals by rationals.--}

\begin{figure}[h]
\includegraphics[width=0.50\textwidth]{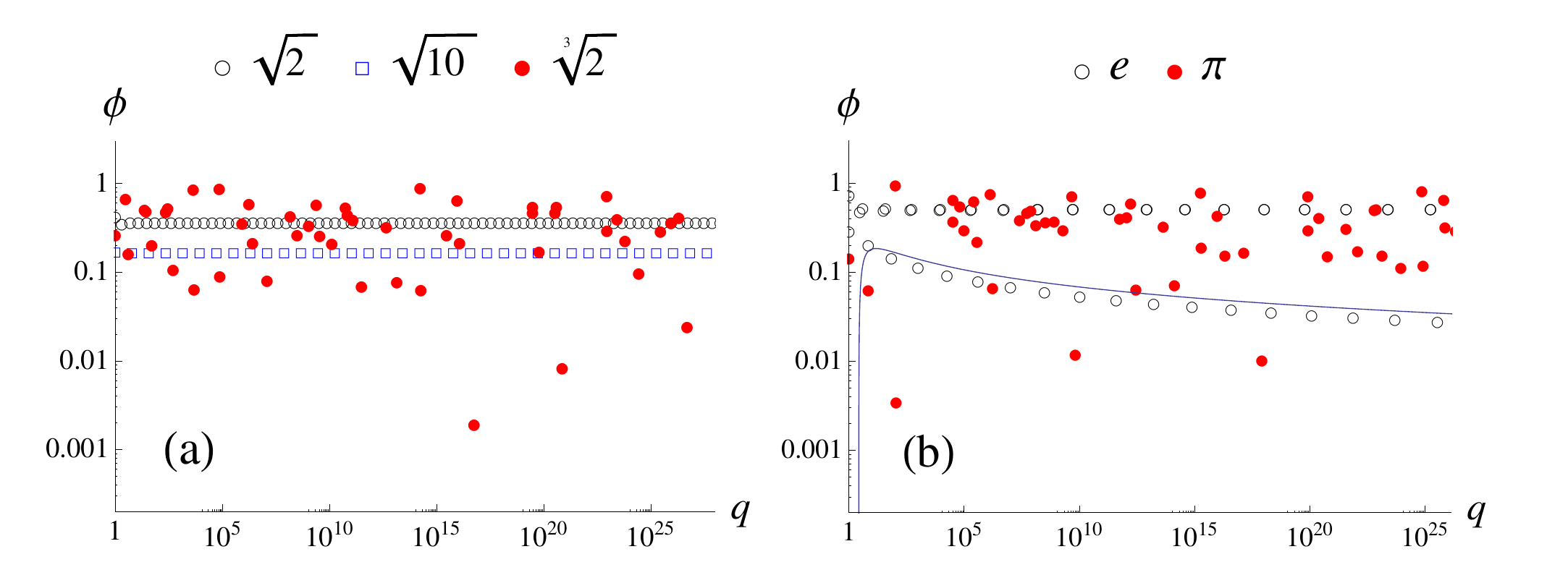}
\caption{Relative error $\phi=|x-p/q|q^2$ of the best rational approximations $p/q$ --the continued fractional approximations-- to several irrational numbers: (a) $x=\sqrt{2}$ (empty circles), $\sqrt{10}$ (empty squares), and $2^{1/3}$ (filled circles); (b) $e$ (empty circles) and $\pi$ (filled circles), as a function of the denominator $q$. The solid line for $e$ illustrates Davis' result with $\phi=\log(\log(q))/\log(q)/2$.
}
\label{fig:numbertheory} 
\end{figure}

Hurwitz's theorem \cite{hardy} states that for any irrational number $x$, there are infinitely many rational approximations $p/q$ such that $\phi=|x-p/q|q^2<1/\sqrt{5}$, providing a universal upper bound to $\phi$. 

Liouville  proved  \cite{hardy} that for every {\it quadratic irrational} $x$ -- i.e. the roots of a quadratic equation with integer coefficients -- there is a positive constant $c_0$ such that $c_0<\phi$ holds for every positive integer $p$ and $q$. Provided this constant is not very small ---which is indeed the case for $x=\sqrt{2}$ and $\sqrt{10}$, as shown in Fig.  \ref{fig:numbertheory}--- the quadratic irrational is difficult to approximate by rational numbers and  quasiperiodic behavior  is guaranteed for arbitrary  $T_s$.

Other irrational numbers admit better rational approximations. For example, 
 Davis proved in 1978 \cite{dav78} that for the transcendental number $e$, for each $\hat{c}>1/2$ there are infinitely many rational approximations with $q\ge2$ that satisfy $\phi \le \hat{c}\log(\log(q))/\log(q)$, 
which is also illustrated in Fig. \ref{fig:numbertheory}. In our non-linear system, Davis' result and (\ref{eq:Tses}) yield periodic behavior for large enough denominators $q$, though the decay is very slow and the corresponding  times $T_s$  over which the effect can be observed are impractically large.
One of the transcendental numbers that admits the best rational approximations is the Liouville constant $\alpha=\sum_{j=1}^\infty 10^{-j!}$, constructed by Liouville to this effect, establishing in this way the existence of a transcendental number for the first time. It will also be considered to  numerically numerically our results.


\begin{figure}
\begin{center}
\includegraphics[width=0.45\textwidth]{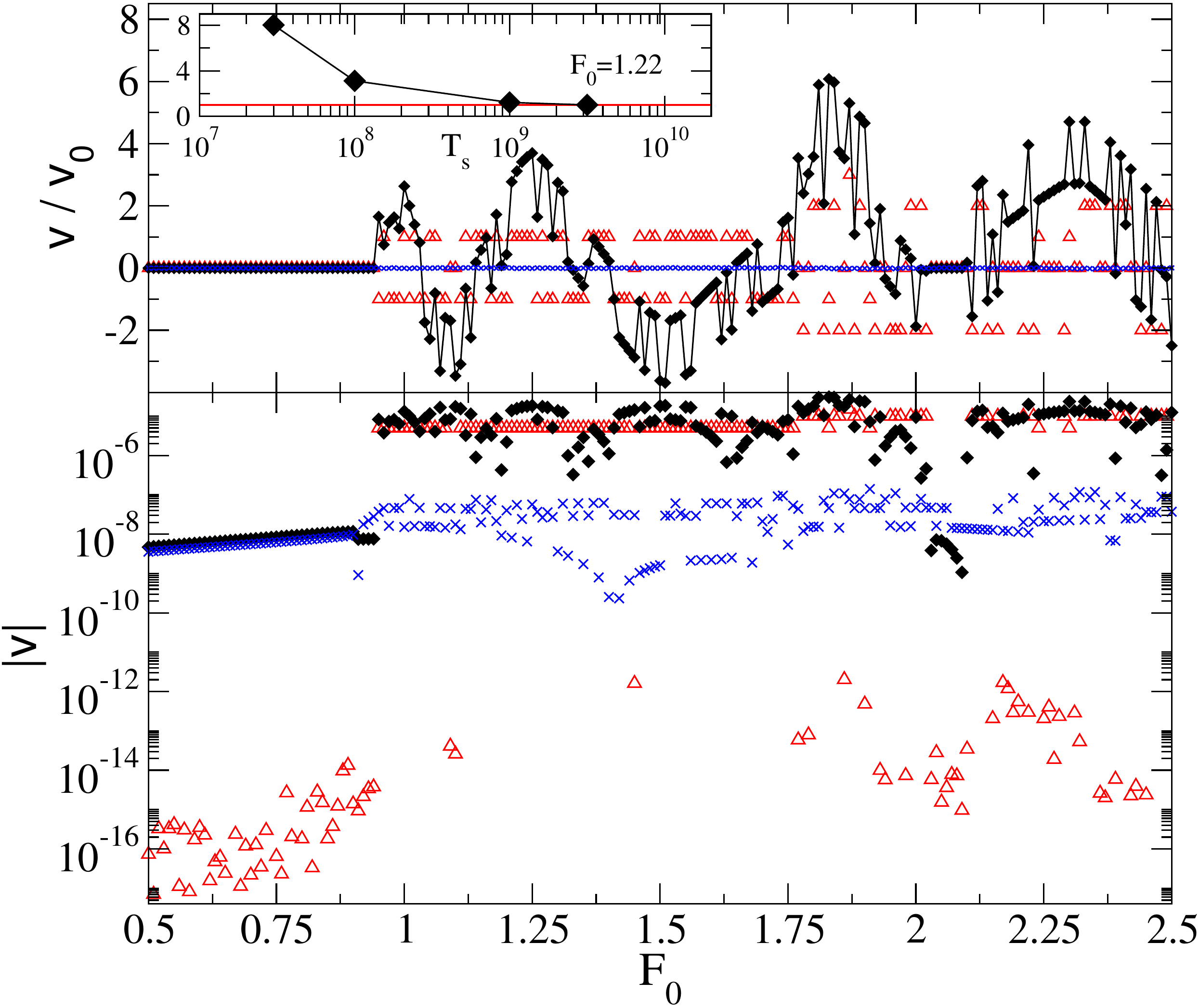}
\end{center}
\caption{(Color online) Current vs. driving amplitude for the driving $F_1$.  Filled (black) diamonds correspond to the ratio $\omega_2/\omega_1=10\alpha=10\sum_{j=1}^\infty 10^{-j!}=1.100010\ldots\,$ and   $T_s=10^8$. Triangles (red) depict the results for the driving with the rational approximation $\omega_2/\omega_1=110001/10^5$, 
and the (blue) crosses to the quadratic irrational  $\omega_2/\omega_1=\sqrt{6/5}=1.095\ldots$ for  $T_s=10^8$. Inset: current as a function of $ T_s$  for 
 $\omega_2/\omega_1 =10\alpha$ and $F_0=1.22$, showing convergence to the 
 rational approximation value (horizontal solid line). The bottom panel shows the top panel data  in a semilogarithm scale.
}
\label{fig:liouville} 
\end{figure}

\begin{figure}
\begin{center}
\includegraphics[width=0.35\textwidth]{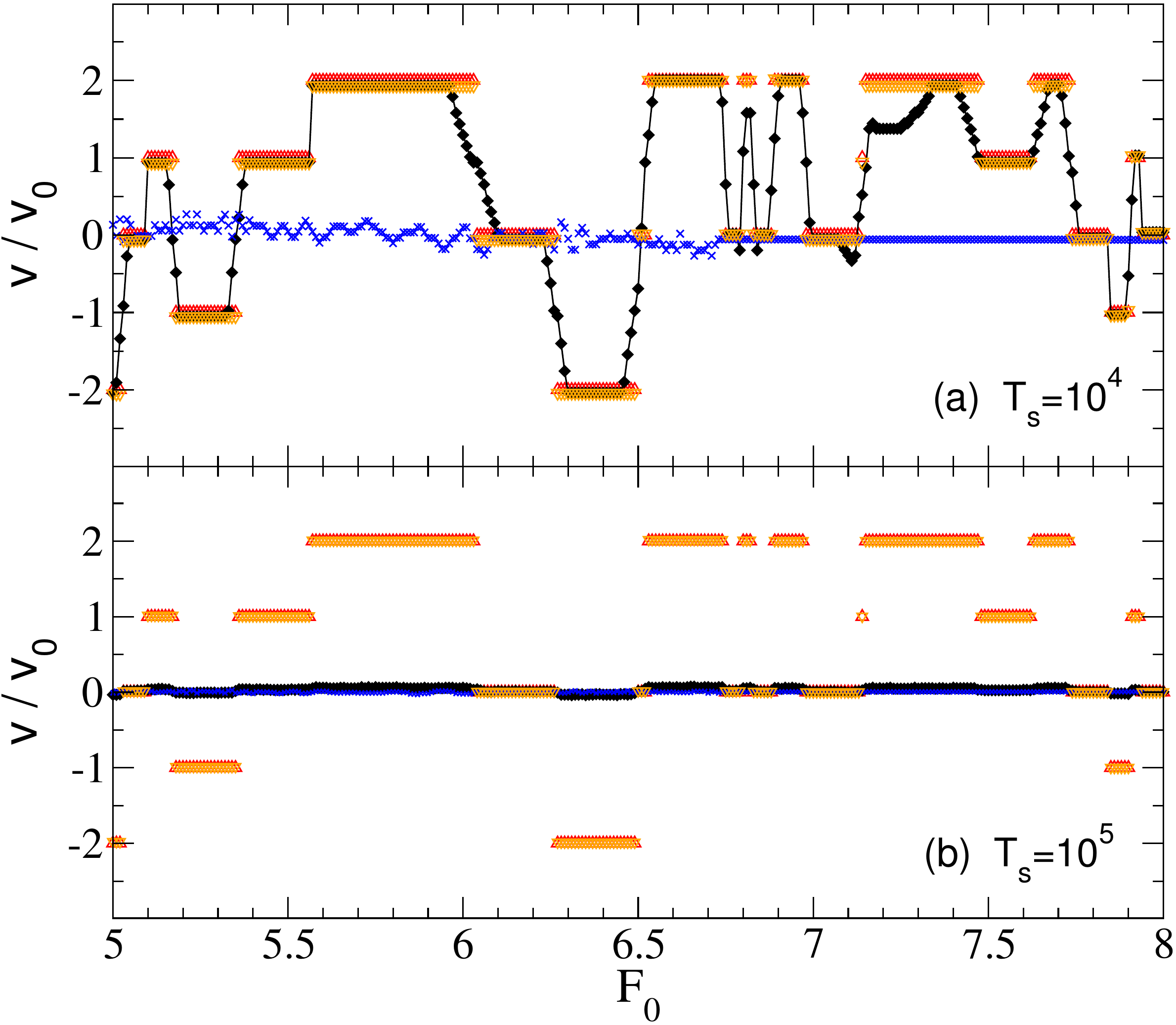}
\end{center}
\caption{(Color online) Current vs. $F_0$ for $\omega_2/\omega_1=\pi$ (filled diamonds) with  $T_s=10^4$ (top panel) and $T_s=10^5$ (bottom panel). 
 Triangles up (red) show the results for the driving with the rational approximation $\omega_2/\omega_1=355/113$, while triangles down (orange) are the results for the same frequency ratio but with the corresponding  $T_s=10^4$ and $10^5$. Crosses (blue) correspond to $\omega_2/\omega_1=\sqrt{10}$. 
}
\label{fig:pi} 
\end{figure}

\paragraph*{Numerical examples.--}

The directed current in the overdamped system (\ref{eq:overdamp}) driven by periodic drivings was numerically computed by using  a multiple of the known period $T$ as sampling time $T_s$, after waiting a brief relaxation time. This procedure allows for accurate estimations of $v_{pq}$ even though $T_s$ is not too large. 

Figure~\ref{fig:liouville} shows results for a driving $F_1$ and three different values of the ratio $\omega_2/\omega_1$: the value $\omega_2/\omega_1=110001/10^5$, a very good rational approximation to the Liouville number $10\alpha$ with a very low error $\phi\sim 10^{-13}$, the Liouville number $10 \alpha$ itself, and the quadratic irrational $\sqrt{6/5}$. 
 We consider first the case of $\omega_2/\omega_1=110001/10^5$ (triangles). 
For low values of $F_0$ no current is expected \cite{hanmar09}, a fact that is shown in the bottom panel of Fig.~\ref{fig:liouville} as a measured current of order $10^{-16}$ due to the use of double precision variables in the computer simulations, which provides 15 to 17 significant decimal digits precision. On the other hand, {\it finite} currents are observed to be a few multiples of $v_0=\lambda/T$, which in our units is of order $1/T$, with $T$ itself being of order $10^5$. 
We turn now to the case of the two irrational numbers. Numerical results have to be compared to the infinite-time-limit genuine quasiperiodic behavior: a {\it zero} current  irrespective of $F_0$. For the irrational number $\omega_2/\omega_1=10\alpha$ , (\ref{eq:crit}) predicts for  $T_s=10^8$ that the finite-time current will deviate from the infinite time limit, and will be of  the same order of magnitude as the results of the periodic approximation.
The precise deviations from the periodic data are due  to the fact that $T_s$ is not an exact multiple of the period $T$, as  shown in the inset  of Fig.~\ref{fig:liouville}. Absence of current is shown numerically to be of order $1/T_s=10^{-8}$. 

 In contrast, genuine quasiperiodic behavior, i.e. a {\it zero} current of order $1/T_s$ irrespective of $F_0$, can be seen in the whole range of driving amplitudes shown in Fig.~\ref{fig:liouville} (crosses in both panels) for the quadratic irrational $\omega_2/\omega_1=\sqrt{6/5}$. These current values are three orders of magnitude smaller than the ones shown for the irrational $10\alpha$. Even larger factors are expected as $T_s$ is increased, since the current will be locked to the order of magnitude provided by this periodic approximation until the next one, which has period $T\sim 10^{18}$, replaces it. 

 The same effect can be observed in other irrational numbers with even poorer rational approximations. 
Figure \ref{fig:pi} shows the results for a driving $F_2$ with a frequency ratio $\omega_2/\omega_1=\pi$. By choosing $\varepsilon_0=0.1$, from (\ref{eq:crit}) the best periodic approximation for observation times in the range $T_s=10^2-10^4$ is given by $\omega_2/\omega_1=355/113$. This is indeed what is observed in the top panel of Fig.~\ref{fig:pi} for  $T_s=10^4$, where the predicted periodic behavior is clearly seen. In contrast, it can also be observed that the currents obtained with the quadratic irrational $\omega_2/\omega_1=\sqrt{10}$ are much smaller, in fact of order $1/T_s$, as expected. But quasiperiodic behavior is also recovered for $\omega_2/\omega_1=\pi$ when  $ T_s$  is increased to $10^5$, as shown in the bottom panel of Fig. \ref{fig:pi}. In this case, the rational approximation 355/133 no longer satisfies (\ref{eq:crit}), and the next one is given by $\omega_2/\omega_1=312689/99532$, with an associated period that is about six times larger than $T_s$. 
This behavior can be directly traced back to the sub-Fourier resolution: As clearly seen in Fig.~\ref{fig:freq_resolution}, the frequency separation $|\pi-355/113|\approx2.7\cdot 10^{-7}$ is within the resolution window $\varepsilon_0\Delta\omega_2$ for $T_s=10^4$, but is outside for $T_s=10^5$ despite being well inside the Fourier width $\Delta\omega_F=2\pi/T_s\approx6.3\cdot10^{-5}$ and the window $\varepsilon_0\Delta\omega_F$.

\paragraph*{Conclusions.--}
We have shown, on the basis of symmetry properties only and thus not relying on the specific details of the dynamics, that the response of a nonlinear system to a bi-frequency drive depends on the interaction time and the nature of the frequency ratio between the driving frequencies. In particular, we showed that an irrational ratio is not sufficient to guarantee quasiperiodic response. The system frequency resolution, being intrinsically sub-Fourier, dictates that quasiperiodicity is only guaranteed for irrational numbers which have bad rational approximations, such as many quadratic irrationals \footnote{Not all quadratic irrationals have bad rational approximations, as illustrated by the example $x=1+\sqrt{2}/10^n$, with $n$ an arbitrarily large integer.} .

\acknowledgements{We acknowledge financial support from the Leverhulme Trust and the Ministerio de Ciencia e Innovaci\'on of Spain FIS2008-02873 (D.C. and J. C.-P.).}


\begin{thebibliography}{10}%
\makeatletter
\providecommand \@ifxundefined [1]{%
 \ifx #1\undefined \expandafter \@firstoftwo
 \else \expandafter \@secondoftwo
\fi
}%
\providecommand \@ifnum [1]{%
 \ifnum #1\expandafter \@firstoftwo
 \else \expandafter \@secondoftwo
\fi
}%
\providecommand \enquote [1]{``#1''}%
\providecommand \bibnamefont  [1]{#1}%
\providecommand \bibfnamefont [1]{#1}%
\providecommand \citenamefont [1]{#1}%
\providecommand\href[0]{\@sanitize\@href}%
\providecommand\@href[1]{\endgroup\@@startlink{#1}\endgroup\@@href}%
\providecommand\@@href[1]{#1\@@endlink}%
\providecommand \@sanitize [0]{\begingroup\catcode`\&12\catcode`\#12\relax}%
\@ifxundefined \pdfoutput {\@firstoftwo}{%
 \@ifnum{\z@=\pdfoutput}{\@firstoftwo}{\@secondoftwo}%
}{%
 \providecommand\@@startlink[1]{\leavevmode}%
 \providecommand\@@endlink[0]{}%
}{%
 \providecommand\@@startlink[1]{%
  \leavevmode
  \pdfstartlink
   attr{/Border[0 0 1 ]/H/I/C[0 1 1]}%
   user{/Subtype/Link/A<</Type/Action/S/URI/URI(#1)>>}%
  \relax
 }%
 \providecommand\@@endlink[0]{\pdfendlink}%
}%
\providecommand \url  [0]{\begingroup\@sanitize \@url }%
\providecommand \@url [1]{\endgroup\@href {#1}{\urlprefix}}%
\providecommand \urlprefix [0]{URL }%
\providecommand \Eprint[0]{\href }%
\@ifxundefined \urlstyle {%
  \providecommand \doi [1]{doi:\discretionary{}{}{}#1}%
}{%
  \providecommand \doi [0]{doi:\discretionary{}{}{}\begingroup
  \urlstyle{rm}\Url }%
}%
\providecommand \doibase [0]{http://dx.doi.org/}%
\providecommand \Doi[1]{\href{\doibase#1}}%
\providecommand \bibAnnote [3]{%
  \BibitemShut{#1}%
  \begin{quotation}\noindent
    \textsc{Key:}\ #2\\\textsc{Annotation:}\ #3%
  \end{quotation}%
}%
\providecommand \bibAnnoteFile [2]{%
  \IfFileExists{#2}{\bibAnnote {#1} {#2} {\input{#2}}}{}%
}%
\providecommand \typeout [0]{\immediate \write \m@ne }%
\providecommand \selectlanguage [0]{\@gobble}%
\providecommand \bibinfo [0]{\@secondoftwo}%
\providecommand \bibfield [0]{\@secondoftwo}%
\providecommand \translation [1]{[#1]}%
\providecommand \BibitemOpen[0]{}%
\providecommand \bibitemStop [0]{}%
\providecommand \bibitemNoStop [0]{.\EOS\space}%
\providecommand \EOS [0]{\spacefactor3000\relax}%
\providecommand \BibitemShut [1]{\csname bibitem#1\endcsname}%
\bibitem{levine}%
  \BibitemOpen
  \bibfield{author}{%
  \bibinfo {author} {\bibfnamefont{D.}~\bibnamefont{Levine}}\ and\ \bibinfo
  {author} {\bibfnamefont{P.~J.}\ \bibnamefont{Steinhardt}},\ }%
  \bibfield{journal}{%
  \bibinfo {journal} {Phys. Rev. Lett.}\ }%
  \textbf{\bibinfo {volume} {53}},\ \bibinfo {pages} {2477} (\bibinfo {year}
  {1984})%
  \bibAnnoteFile{NoStop}{levine}%
\bibitem{guidoni}%
  \BibitemOpen
  \bibfield{author}{%
  \bibinfo {author} {\bibfnamefont{L.}~\bibnamefont{Guidoni}}, \bibinfo
  {author} {\bibfnamefont{C.}~\bibnamefont{Trich\'e}}, \bibinfo {author}
  {\bibfnamefont{P.}~\bibnamefont{Verkerk}},\ and\ \bibinfo {author}
  {\bibfnamefont{G.}~\bibnamefont{Grynberg}},\ }%
  \bibfield{journal}{%
  \bibinfo {journal} {Phys. Rev. Lett.}\ }%
  \textbf{\bibinfo {volume} {79}},\ \bibinfo {pages} {3363} (\bibinfo {year}
  {1997})%
  \bibAnnoteFile{NoStop}{guidoni}%
\bibitem{feudel06}%
  \BibitemOpen
  \bibfield{author}{%
  \bibinfo {author} {\bibfnamefont{U.}~\bibnamefont{Feudel}}, \bibinfo {author}
  {\bibfnamefont{S.}~\bibnamefont{Kuznetsov}},\ and\ \bibinfo {author}
  {\bibfnamefont{A.}~\bibnamefont{Pikovsky}},\ }%
  \emph{\bibinfo {title} {Strange Nonchaotic Attractors, Dynamics between Order
  and Chaos in Quasiperiodically Forced Systems}}\ (\bibinfo {publisher} {World
  Sicentific Publishing},\ \bibinfo {address} {Singapure},\ \bibinfo {year}
  {2006})%
  \bibAnnoteFile{NoStop}{feudel06}%
\bibitem{neupik02}%
  \BibitemOpen
  \bibfield{author}{%
  \bibinfo {author} {\bibfnamefont{E.}~\bibnamefont{Neumann}}\ and\ \bibinfo
  {author} {\bibfnamefont{A.}~\bibnamefont{Pikovsky}},\ }%
  \bibfield{journal}{%
  \bibinfo {journal} {Eur. Phys. J. B}\ }%
  \textbf{\bibinfo {volume} {26}},\ \bibinfo {pages} {219} (\bibinfo {year}
  {2002})%
  \bibAnnoteFile{NoStop}{neupik02}%
\bibitem{gomden06}%
  \BibitemOpen
  \bibfield{author}{%
  \bibinfo {author} {\bibfnamefont{R.}~\bibnamefont{Gommers}}, \bibinfo
  {author} {\bibfnamefont{S.}~\bibnamefont{Denisov}},\ and\ \bibinfo {author}
  {\bibfnamefont{F.}~\bibnamefont{Renzoni}},\ }%
  \bibfield{journal}{%
  \bibinfo {journal} {Phys. Rev. Lett.}\ }%
  \textbf{\bibinfo {volume} {96}},\ \bibinfo {pages} {240604} (\bibinfo {year}
  {2006})%
  \bibAnnoteFile{NoStop}{gomden06}%
\bibitem{cubren12}%
  \BibitemOpen
  \bibfield{author}{%
  \bibinfo {author} {\bibfnamefont{D.}~\bibnamefont{Cubero}}\ and\ \bibinfo
  {author} {\bibfnamefont{F.}~\bibnamefont{Renzoni}},\ }%
  \bibfield{journal}{%
  \bibinfo {journal} {Phys. Rev. E}\ }%
  \textbf{\bibinfo {volume} {86}},\ \bibinfo {pages} {056201} (\bibinfo {year}
  {2012})%
  \bibAnnoteFile{NoStop}{cubren12}%
\bibitem{glalib88}%
  \BibitemOpen
  \bibfield{author}{%
  \bibinfo {author} {\bibfnamefont{J.~A.}\ \bibnamefont{Glazier}}\ and\
  \bibinfo {author} {\bibfnamefont{A.}~\bibnamefont{Libchaber}},\ }%
  \bibfield{journal}{%
  \bibinfo {journal} {IEEE Trans. on circuits and systems}\ }%
  \textbf{\bibinfo {volume} {35}},\ \bibinfo {pages} {790} (\bibinfo {year}
  {1988})%
  \bibAnnoteFile{NoStop}{glalib88}%
\bibitem{sos58}%
  \BibitemOpen
  \bibfield{author}{%
  \bibinfo {author} {\bibfnamefont{V.}~\bibnamefont{Sos}},\ }%
  \bibfield{journal}{%
  \bibinfo {journal} {ann. Univ. Sci. Budapest Eotvos Sect. Math.}\ }%
  \textbf{\bibinfo {volume} {1}},\ \bibinfo {pages} {127} (\bibinfo {year}
  {1958})%
  \bibAnnoteFile{NoStop}{sos58}%
\bibitem{sla67}%
  \BibitemOpen
  \bibfield{author}{%
  \bibinfo {author} {\bibfnamefont{N.}~\bibnamefont{Slater}},\ }%
  \bibfield{journal}{%
  \bibinfo {journal} {Proc. Camb. Phil. Soc.}\ }%
  \textbf{\bibinfo {volume} {63}},\ \bibinfo {pages} {1115} (\bibinfo {year}
  {1967})%
  \bibAnnoteFile{NoStop}{sla67}%
\bibitem{Note1}%
  \BibitemOpen
  \bibinfo {note} {See Supplemental Material at for the explicit validation of
  our approach in the case of a spatially non-periodic system.}%
  \bibAnnoteFile{Stop}{Note1}%
\bibitem{rei01}%
  \BibitemOpen
  \bibfield{author}{%
  \bibinfo {author} {\bibfnamefont{P.}~\bibnamefont{Reimann}},\ }%
  \bibfield{journal}{%
  \bibinfo {journal} {Phys. Rev. Lett.}\ }%
  \textbf{\bibinfo {volume} {86}},\ \bibinfo {pages} {4992} (\bibinfo {year}
  {2001})%
  \bibAnnoteFile{NoStop}{rei01}%
\bibitem{reimann02}%
  \BibitemOpen
  \bibfield{author}{%
  \bibinfo {author} {\bibfnamefont{P.}~\bibnamefont{Reimann}},\ }%
  \bibfield{journal}{%
  \bibinfo {journal} {Phys. Rep.}\ }%
  \textbf{\bibinfo {volume} {361}},\ \bibinfo {pages} {57} (\bibinfo {year}
  {2002})%
  \bibAnnoteFile{NoStop}{reimann02}%
\bibitem{ajdmuk94}%
  \BibitemOpen
  \bibfield{author}{%
  \bibinfo {author} {\bibfnamefont{A.}~\bibnamefont{Ajdari}}, \bibinfo {author}
  {\bibfnamefont{D.}~\bibnamefont{Mukamel}}, \bibinfo {author}
  {\bibfnamefont{L.}~\bibnamefont{Peliti}},\ and\ \bibinfo {author}
  {\bibfnamefont{J.}~\bibnamefont{Prost}},\ }%
  \bibfield{journal}{%
  \bibinfo {journal} {J. Phys. I France}\ }%
  \textbf{\bibinfo {volume} {4}},\ \bibinfo {pages} {1551} (\bibinfo {year}
  {1994})%
  \bibAnnoteFile{NoStop}{ajdmuk94}%
\bibitem{fladen04}%
  \BibitemOpen
  \bibfield{author}{%
  \bibinfo {author} {\bibfnamefont{S.}~\bibnamefont{Flach}}\ and\ \bibinfo
  {author} {\bibfnamefont{S.}~\bibnamefont{Denisov}},\ }%
  \bibfield{journal}{%
  \bibinfo {journal} {Acta Phys. Pol. B}\ }%
  \textbf{\bibinfo {volume} {35}},\ \bibinfo {pages} {1437} (\bibinfo {year}
  {2004})%
  \bibAnnoteFile{NoStop}{fladen04}%
\bibitem{cascub13}%
  \BibitemOpen
  \bibfield{author}{%
  \bibinfo {author} {\bibfnamefont{J.}~\bibnamefont{Casado-Pascual}}, \bibinfo
  {author} {\bibfnamefont{D.}~\bibnamefont{Cubero}},\ and\ \bibinfo {author}
  {\bibfnamefont{F.}~\bibnamefont{Renzoni}},\ }%
  \bibfield{journal}{%
  \bibinfo {journal} {Phys. Rev. E}\ }%
  \textbf{\bibinfo {volume} {88}},\ \bibinfo {pages} {062919} (\bibinfo {year}
  {2013})%
  \bibAnnoteFile{NoStop}{cascub13}%
\bibitem{quicue10}%
  \BibitemOpen
  \bibfield{author}{%
  \bibinfo {author} {\bibfnamefont{N.~R.}\ \bibnamefont{Quintero}}, \bibinfo
  {author} {\bibfnamefont{J.~A.}\ \bibnamefont{Cuesta}},\ and\ \bibinfo
  {author} {\bibfnamefont{R.}~\bibnamefont{Alvarez-Nodarse}},\ }%
  \bibfield{journal}{%
  \bibinfo {journal} {Phys. Rev. E}\ }%
  \textbf{\bibinfo {volume} {81}},\ \bibinfo {pages} {030102R} (\bibinfo {year}
  {2010})%
  \bibAnnoteFile{NoStop}{quicue10}%
\bibitem{cuequi13}%
  \BibitemOpen
  \bibfield{author}{%
  \bibinfo {author} {\bibfnamefont{J.}~\bibnamefont{Cuesta}}, \bibinfo {author}
  {\bibfnamefont{N.}~\bibnamefont{Quintero}},\ and\ \bibinfo {author}
  {\bibfnamefont{R.}~\bibnamefont{Alvarez-Nodarse}},\ }%
  \bibfield{journal}{%
  \bibinfo {journal} {Phys. Rev. X}\ }%
  \textbf{\bibinfo {volume} {3}},\ \bibinfo {pages} {041014} (\bibinfo {year}
  {2013})%
  \bibAnnoteFile{NoStop}{cuequi13}%
\bibitem{hardy}%
  \BibitemOpen
  \bibfield{author}{%
  \bibinfo {author} {\bibfnamefont{G.~H.}\ \bibnamefont{Hardy}}\ and\ \bibinfo
  {author} {\bibfnamefont{E.~M.}\ \bibnamefont{Wright}},\ }%
  \emph{\bibinfo {title} {An introduction to the theory of numbers}}\ (\bibinfo
  {publisher} {Oxford University Press},\ \bibinfo {address} {London},\
  \bibinfo {year} {1975})%
  \bibAnnoteFile{NoStop}{hardy}%
\bibitem{dav78}%
  \BibitemOpen
  \bibfield{author}{%
  \bibinfo {author} {\bibfnamefont{C.~S.}\ \bibnamefont{Davis}},\ }%
  \bibfield{journal}{%
  \bibinfo {journal} {J. Austral. Math. Soc. Ser. A}\ }%
  \textbf{\bibinfo {volume} {25}},\ \bibinfo {pages} {407} (\bibinfo {year}
  {1978})%
  \bibAnnoteFile{NoStop}{dav78}%
\bibitem{hanmar09}%
  \BibitemOpen
  \bibfield{author}{%
  \bibinfo {author} {\bibfnamefont{P.}~\bibnamefont{H\"anggi}}\ and\ \bibinfo
  {author} {\bibfnamefont{F.}~\bibnamefont{Marchesoni}},\ }%
  \bibfield{journal}{%
  \bibinfo {journal} {Rev. Mod. Phys.}\ }%
  \textbf{\bibinfo {volume} {81}},\ \bibinfo {pages} {387} (\bibinfo {year}
  {2009})%
  \bibAnnoteFile{NoStop}{hanmar09}%
\bibitem{Note2}%
  \BibitemOpen
  \bibinfo {note} {Not all quadratic irrationals have bad rational
  approximations, as illustrated by the example $x=1+\protect \sqrt {2}/10^n$,
  with $n$ an arbitrarily large integer.}%
  \bibAnnoteFile{Stop}{Note2}%
\end{thebibliography}
%

\end{document}